\documentstyle[preprint,aps,epsf,floats]{revtex}

\begin{document}

\preprint{\tighten \vbox{\hbox{} }}

\title{Extracting $V_{ub}$ Without Recourse to Structure Functions}

\author{Adam K.\ Leibovich, Ian Low, and I.\ Z.\ Rothstein}

\address{
Department of Physics,
Carnegie Mellon University,
Pittsburgh, PA 15213}

\maketitle

{\tighten
\begin{abstract}

We present a closed form expression for $|V_{ub}|^2/ |V_{tb}
V_{ts}^*|^2$ in terms of the endpoint photon and lepton spectra from
the inclusive decays $B\rightarrow X_s\,\gamma$ and $B\rightarrow
X_u\,\ell\,\nu$, respectively, which includes the resummation of the
endpoint logs at next to leading order and is completely independent
of the $B$ meson structure function.  The use of this expression for
extracting $V_{ub}$ would eliminate the large systematic errors
usually incurred due to the modeling of the heavy quarks' Fermi
motion.

\end{abstract}
}


\newpage

\section{Introduction}

Presently great effort is being put into determining the nature of CP
violation in the Standard Model. The unitarity triangle is being
constrained with increasing precision via a combination of
measurements. The size of the allowed regions in the $(\rho,\eta)$
plane \cite{BA} could be greatly reduced if we could more precisely
calculate the hadronic dynamics involved in weak decays.  Indeed, one
of the largest uncertainties in determining the position of the upper
vertex of the unitarity triangle stems from our poor knowledge of the
CKM element $V_{ub}$.

Extracting $V_{ub}$ presents quite a challenge to the theoretical
community. The elegant methods of heavy quark effective theory (HQET)
\cite{MW} used in determining $V_{cb}$ from exclusive $B$ to $D$
transitions are of little use in the heavy to light modes
\cite{ligeti}. Whereas the extraction from inclusive decays is
hindered by the experimental cuts which force theorists to make
predictions in a region of phase space which is especially sensitive
to non-perturbative hadronic dynamics.  The presence of these
experimental cuts thus introduces theoretical errors, due to the use
of models, which are difficult, if not impossible, to quantify.  Here
we will present a method for extracting $|V_{ub}|$ which has the
benefit of avoiding the issue of the unknown hadronic dynamics, thus
obviating the need for any models.

To extract $|V_{ub}|$ from inclusive semi-leptonic $B$ decays, we may
study the endpoint region of the charged lepton energy spectrum in
$B\rightarrow X_u \ell \bar{\nu}$ decays and apply a lower cut on
$E_\ell$ above the $b\rightarrow c$ endpoint energy in order to
eliminate the background from charmed decays.  In the Standard Model,
this decay rate can be calculated in a systematic expansion in
$\alpha_s$, $1/(M_W,m_t)$ and $1/m_b$. Large amounts of resources have
gone into calculating the decay rate at next to leading order in the
strong coupling.  The calculation is broken into three stages. The
full standard model is first matched onto the four fermion theory at
the scale $M_W$. It is then run down to the scale $m_b$, after which
the inclusive decay rate may be calculated in HQET \cite{CGG,BSUV} in
expansion in $1/m_b$.

However, near the endpoint region certain effects arise which
jeopardize the accuracy of the calculation. The expansion in
$\alpha_s$ breaks down in the endpoint region due to presence of the
threshold logs, $\log(1-2 E_\ell / m_b)$.  In addition, the expansion
in $1/m_b$ breaks down because the relevant expansion parameter in
this region is $1/(m_b-2E_\ell) \equiv 1/[m_b(1-x)]$ and the true
kinematic endpoint is determined by the hadronic mass $M_B$, instead
of the partonic mass $m_b$. Thus, both perturbative and
non-perturbative expansions break down at the endpoint. Given the
proximity of the $b\to u$ and $b\to c$ endpoints, there is only a very
narrow window, $\Delta E \approx 350\,\mbox{MeV}$, left for the
extraction of $|V_{ub}|$.  In particular, due to the fact that there
is an imposed energy cut on the charged lepton, the small parameter
$1-2E_\ell/m_b$ entering the calculation hinders our ability to make
reliable theoretical predictions. Both the perturbative expansion in
$\alpha_s$ as well as the non-perturbative expansion in $\Lambda_{\rm
QCD}/m_b$ must be reorganized, in a systematic expansion in powers of
$\alpha_s \log(1-x)$ and $\Lambda_{\rm QCD}/[m_b(1-x)]$, respectively.
While the perturbative expansion is, in principle, calculable, the
reorganization of the non-perturbative expansion necessitates the
introduction of a non-perturbative structure function, of which we are
ignorant. Thus, it would seem that we are left in a lurch as far as an
extraction of $|V_{ub}|$ from inclusive decays is concerned.  

However, as is well known \cite{N,MN}, the aforementioned structure
function is universal in $B$ decays, in that it is independent of the
short distance decay under consideration.  Therefore, we may extract
it from another decay, such as $B\rightarrow X_s\gamma$ for use in the
prediction for $B\rightarrow X_u \ell \bar{\nu}$. However, given the
dearth of data near the endpoint this extraction will incur large
errors, given that we don't know its shape. It would be preferable if
we could sidestep the extraction completely.  In this paper we will
show that the process of extraction of the structure function may
indeed be bypassed, and that a prediction for the ratio
$|V_{ub}|^2/|V_{ts}^* V_{tb}|^2$ may be given at next to leading order
in $\log(1-x)$ without model dependence.

\section{The one loop result}

The calculation of inclusive semi-leptonic $B\rightarrow X_u \ell
\bar{\nu}$ decays begins with the low-energy effective Hamiltonian
\cite{buras}
\begin{eqnarray}
H_{eff} &=& -\frac{4G_F}{\sqrt{2}} V_{ub}
             (\bar{u}\gamma_\mu P_L b)\ 
             (\bar{\ell}\gamma^\mu P_L \nu_\ell)  \nonumber\\
        &=& -\frac{4G_F}{\sqrt{2}}V_{ub}\ J_\mu\ J_\ell^{\mu},
\end{eqnarray}
where $G_F$ is the Fermi constant and $P_L$ is the left-handed
projection operator $\frac{1}{2}(1-\gamma_5)$.  $J^\mu$ and
$J^\mu_\ell$ are the hadronic and leptonic currents, respectively.
The differential decay distribution is
\begin{equation}
\frac{d^3\Gamma}{dq^2 dE_\ell dE_\nu} = 2 G_F^2 |V_{ub}|^2
     W_{\alpha\beta}L^{\alpha\beta},
\end{equation}
where $q^2=(p_\ell+p_\nu)^2$ and the leptonic tensor is
\begin{equation}
L^{\alpha\beta}=2\,(p^\alpha_\ell p^\beta_\nu +p^\beta_\ell p^\alpha_\nu
   -g^{\alpha\beta}p_\ell\cdot p_\nu
   -i \epsilon^{\eta\beta\lambda\alpha}p_{\ell\eta}p_{\nu\lambda}).
\end{equation}
The hadronic tensor is related to the imaginary part of a time ordered
product of hadronic currents
\begin{equation}
W_{\alpha\beta} = - \frac{1}{\pi} \mbox{Im}\ T_{\alpha\beta},
\end{equation}
\begin{equation}
T_{\alpha\beta}= -\frac{i}{2M_B}\int d^4 x\ e^{-i \,q\cdot x}
   \langle B|T(J_{\alpha}^\dagger(x)J_{\beta}(0))|B\rangle.
\end{equation}
It is this time ordered product which can be computed using an
operator product expansion (OPE) in terms of local operators,
involving $b$-quark fields in HQET, whose Wilson coefficients can be
calculated in perturbation theory away from the cuts of
$T_{\alpha\beta}$ \cite{CGG,BSUV}.  In particular, the one loop decay
spectrum, including the leading non-perturbative corrections, is given
by \cite{ONEL}
\begin{eqnarray}
\label{oneloop}
\frac{1}{\Gamma_0} \frac{d\Gamma}{dx}
 &=& \theta(1-x)\left[\frac{5x^3}{3}\frac{\lambda_1}{m_b^2}+(6+5x)x^2
     \frac{\lambda_2}{m_b^2}\right]
     -\frac{\lambda_1+33\lambda_2}{6m_b^2}\delta(1-x)
     -\frac{\lambda_1}{6m_b^2}\delta'(1-x) \nonumber\\
  &&+ \theta(1-x)x^2(3-2x)\left[1-\frac{2\alpha_s}{3\pi} G(x)\right],
\end{eqnarray}
where
\begin{eqnarray}
G(x) &=& \log^2(1-x)+2 {\rm Li}_2(x)+\frac{2\pi^2}{3}+
         \frac{82-153x+86x^2}{12x(3-2x)}\nonumber\\
     & & +\frac{41-36x+42x^2-16x^3}{6x^2(3-2x)}\log(1-x) \nonumber\\
     &=& \log^2(1-x)+\frac{31}{6}\log(1-x)+\pi^2+\frac{5}{4}
	+{\mathcal O}(1-x) \qquad {\mathrm as\ }x \to 1,\\
\Gamma_0 &=& |V_{ub}|^2 \frac{G_F^2 m_b^5}{96\pi^3},
\end{eqnarray}
and $x = 2 E_\ell/m_b$.  In Eq.~(\ref{oneloop}) the first line is the
non-perturbative corrections to order $(\Lambda_{\rm QCD}/m_b)^2$ and
the second line contains the perturbative corrections to order
$\alpha_s$. The non-perturbative corrections to this order are
characterized by only two hadronic matrix elements
\begin{eqnarray}
\lambda_1 &=&
  \langle B(v)|\,\bar{b}_v (i D)^2 b_v\,|B(v)\rangle ,\\
\lambda_2 &=& \frac{1}{3}
\langle B(v)|\,\bar{b}_v\frac{g}{2}\sigma_{\mu\nu}G^{\mu\nu}b_v\,|B(v)\rangle.
\end{eqnarray}
Here $b_v$ is the velocity dependent $b$-quark field as defined in
HQET, $g$ is the strong coupling, and $G^{\mu\nu}$ is the strong
interaction field strength tensor. $\lambda_2$ can be determined from
the $B-B^*$ mass difference, yielding $\lambda_2 \approx
0.12\,\mbox{GeV}^2$. From the above expression we see that the
breakdown of the expansions, in $\alpha_s$ and $\Lambda_{\rm
QCD}/m_b$, manifest themselves in the large logs and the derivative of
delta functions, respectively. Though the differential decay rate is
singular as $x\to 1$, these endpoint singularities are integrable and
the total decay rate is given by
\begin{equation}
\Gamma = \frac{\Gamma_0}2\left[1+\frac{\lambda_1-9\lambda_2}{2m_b^2}
         -\frac{2\alpha_s}{3\pi}\left(\pi^2-\frac{25}{4}\right)\right].
\end{equation}

The maximum lepton energy for a particular hadronic final state $X$ is
$E_\ell = (M_B^2-M_X^2)/(2M_B)$ in the $B$ rest frame, and consequently
leptons with energies in the endpoint region (neglecting the pion
mass) $M_B/2 > E_\ell > (M_B^2-M_D^2)/(2M_B)$ can arise only from $b \to
u$ transition. This translates into a lepton energy cut $x^c\approx
0.87$.

\section{The Perturbative Resummation}

The systematics of the perturbative expansion is discussed at length
in Ref.~\cite{LR}. Here we summarize the discussion.  The perturbative
series near the endpoint schematically takes the form
\begin{eqnarray}
 && C_{00} \nonumber\\*
&+& C_{12} \alpha_s\log^2(1-x)
  + C_{11} \alpha_s\log(1-x) + C_{10} \alpha_s \nonumber\\
&+& C_{24} \alpha_s^2\log^4(1-x)
  + C_{23} \alpha_s^2\log^3(1-x)+ C_{22} \alpha_s^2\log^2(1-x)
  + C_{21} \alpha_s^2\log(1-x) + C_{20} \alpha_s^2 \nonumber\\
&+& C_{36} \alpha_s^3\log^6(1-x)
  + C_{35} \alpha_s^3\log^5(1-x)
  + C_{34} \alpha_s^3\log^4(1-x) + C_{33} \alpha_s^3\log^3(1-x)
  + \cdots \nonumber\\*
&+& \cdots.
\end{eqnarray}
Naively, if we would like to go to a region where
$\alpha_s^m\log^n(1-x) \sim 1$, it seems that we would have to sum a
large number of subleading logarithms below the line $C_{mk,nk}$, a
daunting task.

However, this criteria is not correct. The reason for this is that the
series is known to exponentiate into a particular form. In fact, the
series resums into a function of the form
\begin{eqnarray}
\label{sudseries}
\log\left[\frac{d\Gamma}{dx}\right] 
   &\sim& \log(1-x) g_1[\alpha_s \beta_0 \log(1-x)]
   +g_2[\alpha_s \beta_0 \log(1-x)] \nonumber\\
   & & + \alpha_s g_3[\alpha_s \beta_0 \log(1-x)]+\cdots,
\end{eqnarray}
where $\beta_0=(33-2 N_f)/(12 \pi)$.  This result implies that the
aforementioned triangle assumes a definite structure. The resummation
does {\it not} sum the entire triangle, but the terms dropped are
subleading in the exponent.  For more details see \cite{LR}.

The resummation itself is performed in moment space, where the rate
factorizes into a short distance hard part $(H_N)$, a soft part
$(\sigma_N)$, and a jet function $(J_N)$ \cite{SK}.  Using
renormalization group techniques it is possible to show that \cite{AR}
\begin{equation}
\label{resummed}
\sigma_NJ_N = \exp\left[\log N g_1(\chi) + g_2(\chi)\right],
\end{equation}
where $\chi = \alpha_s(m_b^2)\beta_0\log N$,
\begin{eqnarray}
\label{g1}
g_1 &=& -\frac2{3\pi\beta_0\chi}
      [(1-2\chi)\log(1-2\chi) - 2(1-\chi)\log(1-\chi)], \\
\label{g2}
g_2 &=& -\frac{k}{3\pi^2\beta_0^2}
            \left[2\log(1-\chi)-\log(1-2\chi)\right]
   -\frac{2\beta_1}{3\pi\beta_0^3}\left[\log(1-2\chi)-2\log(1-\chi)
\frac{}{}\right.\nonumber\\
&& + \left.\,\frac12\log^2(1-2\chi)-\log^2(1-\chi)\right]
   -\frac1{\pi\beta_0}\log(1-\chi)
   -\frac2{3\pi\beta_0}\log(1-2\chi) \nonumber\\
&& + \, \frac{4\gamma_E}{3\pi\beta_0}
    \left[\log(1-2\chi) - \log(1-\chi)\right] + g_{s\ell}(\chi,x_\nu),
\end{eqnarray}
and
\begin{equation}
g_{s\ell}(\chi,x_\nu) = \frac4{3\pi \beta_0} \log(1-x_\nu) \log(1-\chi).
\end{equation}
In the above, $\beta_1 = (153-19\,N_f)/(24\pi^2)$, $k = 67/6-\pi^2/2 -
5\,N_f/9$, $x_\nu = 2E_\nu / m_b$, and $\gamma_E = 0.577216...$ is the
Euler-Mascheroni constant.  The moment space resummation can be
written as \cite{AR}
\begin{eqnarray}
\label{moment}
M_N &=& -\frac1{\Gamma_0} \int_0^{M_B/m_b}dx\, x^{N-1}\, 
            \frac{d}{dx} \frac{d\Gamma}{dx} \nonumber\\
        &=& S_N \int_0^1 dx_\nu\, 6 (1-x_\nu) x_\nu
   H(x_\nu) e^{\log N\, g_1(\chi) +g_2(\chi)}
 \; + \; {\cal O} \left(\frac1N\right).
\end{eqnarray}
Note that in writing the above we have taken into account moments of
the non-perturbative structure function, $S_N$, which arises from
resumming leading twist terms in non-perturbative expansion in $1/m_b$
(see \cite{AR} for details.) The hard part is given by\footnote{The
hard part given here is different from that in Ref.~\cite{AR}.  In the
Appendix we perform the inverse Mellin transform to next-to-leading
log accuracy, therefore the hard part can differ since it is a
next-to-next-to-leading contribution. The hard part here is chosen to
reproduce the one loop rate when $x \to 1$ after performing the
inverse Mellin transform given in the Appendix.}
\begin{equation}
H(x_\nu) = 1 - \frac{2\alpha_s}{3\pi}\left[ \frac{2\pi^2}3 + \frac52
    + \log^2(1-x_\nu)-\frac{3}{2}\log(1-x_\nu)
    + \frac{\log(1-x_\nu)}{x_\nu} + 2 {\rm Li}_2(x_\nu) \right].
\end{equation}

To go back to $x$-space, we must take the inverse Mellin transform of
Eq.~(\ref{moment}) which is given by
\begin{eqnarray}
\label{xresum}
\frac{1}{\Gamma_0}\frac{d\Gamma}{dx} &=&
   \frac1{2\pi i}\int_{C-i\infty}^{C+i\infty}M_{N+1} x^{-N}\frac{dN}{N}
\nonumber\\
&=& \int_0^1 dx_\nu\, 6 (1-x_\nu) x_\nu H(x_\nu)
\frac1{2\pi i}\int_{C-i\infty}^{C+i\infty}\,\frac{dN}{N}\,x^{-N}\,S_{N+1}\,
  \sigma_{N+1}\,J_{N+1}.
\end{eqnarray}
Then using the identity derived in the Appendix, and ignoring $S_N$,
we obtain
\begin{eqnarray}
\label{invmell}
&& \frac{1}{2\pi i} \int_{C-i\infty}^{C+i\infty}\,\frac{dN}{N}\,x^{-N}\,
  \exp\{\log(N+1)\,g_1[\alpha_s \beta_0 \log(N+1)]+
  g_2[\alpha_s \beta_0 \log(N+1)]\} \nonumber\\
&&
\phantom{
\frac{1}{2\pi i}\int_{C-i\infty}^{C+i\infty},\frac{dN}{N}\,x^{-N}}
 = \frac{ e^{l\:g_1(\alpha_s \beta_0 l) + g_2(\alpha_s \beta_0 l)}}
    {\Gamma\left[1-g_1(\alpha_s \beta_0 l)-
   \alpha_s \beta_0 l\, g_1'(\alpha_s \beta_0 l)\right]}
   \times \left[1 + {\cal F}(\alpha_s ,l) \right],
\end{eqnarray}
where
\begin{equation}
 {\cal F}(\alpha_s ,l) = \sum_{k=1}^{\infty} \alpha_s^k \,
                      \sum_{j=0}^{k-1} f_{kj}\,l^j
\end{equation}
represents next-to-next-to-leading log contributions and
\begin{eqnarray}
l &=& -\log\left(-\log x\right) \nonumber\\
  &\approx& -\log(1-x) \qquad {\mathrm when\ }x \approx 1.
\end{eqnarray}

It is important at this point to note a crucial difference between
this calculation and resummations carried out in threshold production,
which has been observed in $B \to X_s\gamma$ decays as well.  In
threshold production (Drell-Yan for instance), the overall sign of
$g_1$ is flipped (in standard schemes). This difference in sign
changes the nature of the inverse Mellin transform. In the case of a
negative exponent (our case), the inverse Mellin transform is
integrable, when expanding $g_1$ and $g_2$ in the Sudakov exponent,
whereas it is not the case for threshold production. The expansion of
the series for the Drell-Yan case at leading logarithms accuracy has
introduced factorial growth terms at large order, which shows up as a
non-integrable pole on the positive axis of the Borel plane.  If we
were to perform the inverse Mellin transform exactly, this spurious
factorial growth would be absent, as was pointed out by Catani {\it et
al.}~\cite{CMNT}. It would be interesting to apply the analytic
formulae derived in the Appendix to study the behavior of these
factorially growing terms.

We may use these results to pick out certain terms in the two loop
calculation.  In particular, we may expand the analytic expression and
determine the coefficients of the terms of order
$\alpha_s^2\log^4(1-x)$ and $\alpha_s^2\log^3(1-x)$. Lower order logs
will not be reproduced correctly given the fact that we have dropped
terms of order $\alpha_s$ in the exponent.  Expanding we find
\begin{equation}
\label{BLM}
\left.\frac1{\Gamma_0}
\frac{d\Gamma}{dx}\right|_{{\cal O}(\alpha_s^2)} =
   \frac{2\alpha_s^2}{9\pi^2}\log^4(1-x)+\frac{\alpha_s^2}{\pi^2}
   \left(\frac2{3}\beta_0\pi + \frac{62}{27}\right)\log^3(1-x) +
   \cdots.
\end{equation}
The term proportional to $\alpha_s^2\beta_0$ is part of the BLM
correction, which is typically a good approximation to the full two
loop result. However, note that the contribution of the last term in
Eq.~(\ref{BLM}), not proportional to $\beta_0$, is actually larger
than the corresponding BLM correction, the second term in
Eq.~(\ref{BLM}). Similar situation has been observed in $B \to
X_s\gamma$ decays in Ref.~\cite{LR}. Although one can not rule out the
possibility of a cancellation with other terms such that the BLM
result in the two loop result still dominates, this does cast doubt on
the issue of BLM dominance in the endpoint region, since we know that
Eq.~({\ref{BLM}) is the leading contribution at ${\cal O}(\alpha_s^2)$
when $x \to 1$. Terms we neglect in Eq.~(\ref{BLM}) are suppressed by
at least a factor of $\log(1-x)$ in the endpoint region.

\section{$B \to X_{\lowercase{s}}\gamma$ Decays}

Here we briefly overview the inclusive $B \to X_s\gamma$ decays
\cite{AG} and give the analytic expression for the resummed rate. For
$B \to X_s\gamma$ decays, only the high energy part of the photon
spectrum is experimentally accessible due to large background
cuts. Currently the experimental cut on the photon energy is given by
$E^{cut}_\gamma>2.1{\rm\ GeV}$ \cite{CLEO}.  This introduces a small
dimensionless parameter $1-2E^{cut}_\gamma/m_b$ into the calculation,
resulting in large threshold logs, $\log(1-2E^{cut}_\gamma/m_b)$,
which grow parametrically as $E^{cut}_\gamma$ approaches the endpoint.

The calculation of inclusive decay rates begins with the low-energy
effective Hamiltonian
\begin{equation}
H_{eff} = -\frac{4G_F}{\sqrt{2}}V_{ts}^*V_{tb}
   \sum_{i=1}^8C_i(\mu)O_i(\mu),
\end{equation}
where $C_i(\mu)$ are Wilson coefficients evaluated at a subtraction
point $\mu$, and $O_i(\mu)$ are dimension six operators.  For $B\to
X_s\gamma$, the only operators that give a relevant contribution are
\begin{eqnarray}
O_2 &=& (\bar{c}_{L\alpha}\gamma^\mu b_{L\alpha})
          (\bar{s}_{L\beta}\gamma_\mu c_{L\beta}), \nonumber\\
O_7 &=& \frac{e}{16\pi^2}m_b
      \bar{s}_{L\alpha}\sigma_{\mu\nu}F^{\mu\nu}b_{R\alpha}, \\
O_8 &=& \frac{g}{16\pi^2}m_b \bar{s}_{L\alpha}
   \sigma_{\mu\nu}G_a^{\mu\nu}T^a_{\alpha\beta}b_{R\beta}. \nonumber
\end{eqnarray}

Near the endpoint, the decay rate is dominated by the $O_7$ operator.
In particular, the decay rate due to the $O_7$ operator is
\begin{equation}
\label{onelooprate}
\left. \frac1{\Gamma^\gamma_0}\frac{d\Gamma^\gamma}{dx}\right|_{x<1} = 
   \frac{\alpha_s}\pi \frac{(2x^2-3x-6)x + 2(x^2-3)\log(1-x)}{3(1-x)},
\end{equation}
where $x=2E_\gamma/m_b$ and 
\begin{equation}
\Gamma^\gamma_0 
  = \frac{G_F^2|V_{ts}^*V_{tb}|^2\alpha\,C_7^2\,m_b^5}{32\pi^4}.
\end{equation}
As $x\to1$ this contribution diverges as $\log(1-x)/(1-x)$.  The
integrated rate is finite due to virtual corrections at $x=1$.  The
differential rate in moment space factorizes into the form \cite{SK}
\begin{equation}
\label{srresum}
\sigma_NJ_N^\gamma = \exp\left[\log N g_1(\chi) + g_2^\gamma(\chi)\right],
\end{equation}
where $g_1(\chi)$ is the same as in the semi-leptonic decays
Eq.~(\ref{g1}) and $g_2^\gamma(\chi) = g_2(\chi) -
g_{s\ell}(\chi,\,x_\nu)$ \cite{AR}. $g_2(\chi)$ and
$g_{s\ell}(\chi,\,x_\nu)$ are defined in the semi-leptonic decays
Eq.~(\ref{g2}) and we see that the only difference between these two
cases is that in the semi-leptonic decays we have an additional
$g_{s\ell}$ piece in the Sudakov exponent and the extra $x_\nu$
integration. The moment space resummation is written as \cite{AR}
\begin{eqnarray}
\label{srmoment}
M^\gamma_N &=& \frac1{\Gamma^\gamma_0} \int_0^{M_B/m_b} dx\, x^{N-1}
  \, \frac{d\Gamma^\gamma}{dx} \nonumber\\
 &=& S_N\, H^\gamma \,e^{\log N g_1(\chi) + g_2^\gamma(\chi)},
\end{eqnarray}
and the hard part is given by
\begin{equation}
H^\gamma = 1 - \frac{2\alpha_s}{3\pi}
  \left(\frac{13}{2}+\frac{2\pi^2}{3}\right).
\end{equation}
It is important to note here that the non-perturbative structure
function $S_N$ is universal in $B$ decays and therefore the same as in
the inclusive semi-leptonic decays.  Evaluating the inverse Mellin
transform using Eq.~(\ref{srinvmell}) and ignoring $S_N$, we get
\begin{eqnarray}
\label{bsgaminv}
\frac1{\Gamma^\gamma_0}\frac{d\Gamma^\gamma}{dx} &=&
   \frac1{2\pi i} \int_{C-i\infty}^{C+i\infty}
    M^\gamma_N \, x^{-N}\,dN \nonumber\\
 &=& -x \frac{d}{dx}\left\{
     \theta(1-x)H^\gamma\frac{e^{\,l\,g_1(\alpha_s \beta_0 l)+
     g_2^\gamma(\alpha_s \beta_0 l)}}
   {\Gamma\left[1-g_1(\alpha_s \beta_0 l)-\alpha_s \beta_0 l
     g_1'(\alpha_s \beta_0 l)\right]} \right\}.
\end{eqnarray}

\section{The Non-perturbative Resummation}
\label{CWTSF}

Before we can confront theoretical calculations presented so far with
experimental data, we have to take into account non-perturbative
corrections.  As is well known, as one probes the spectrum closer the
end point, the OPE breaks down, and the leading twist non-perturbative
corrections must be resummed into the $B$ meson structure function
\cite{N,MN}. Formally we may write the light cone distribution
function for the heavy quark inside the meson as
\begin{equation}
f(k_+)=\langle B(v)\mid \bar{b}_v \delta(k_+-iD_+) b_v \mid B(v)\rangle.
\end{equation}
While the shape of this function is unknown,
the first few moments of $f(k_+)$,
\begin{eqnarray}
A_n &=& \int dk_+ k_+^n f(k_+) \nonumber\\
&=& \langle B(v)\mid \bar{b}_v (iD_+)^n b_v \mid B(v)\rangle,
\end{eqnarray}
are known: $A_0=1$, $A_1=0$, and $A_2=-\lambda_1/3$.  $f(k_+)$ has
support over the range $-\infty<k_+<\bar{\Lambda}$, where
$\bar{\Lambda}$ is a universal parameter in HQET which sets the
deviation of $b$-quark mass from $B$-meson mass in the infinite mass
limit.  The values of $\bar{\Lambda}$ and $\lambda_1$ are not
calculable analytically and have to be extracted from experiments or
calculated on the lattice.  A recent analysis \cite{GKLW} gives
$(\bar{\Lambda},\lambda_1)=(0.39\pm0.11\, \mbox{GeV},
-0.19\mp0.10\,\mbox{GeV}^2)$.

The effects of the Fermi motion of the heavy quark can be included by
convoluting the above structure function with the differential rate
\cite{N,MN,DSU},
\begin{equation}
\label{tree}
\frac{d\Gamma}{dE} = \int_{2E-m_b}^{\bar\Lambda}
dk_+ f(k_+) \frac{d\Gamma_p}{dE}(m_b^*),
\end{equation}
where $E$ is the charged lepton energy in the case of semi-leptonic $b
\to u$ decays or the photon energy in the case of $b \to s \gamma$
decays, and $d\Gamma_p/dE$ is the rate from one loop calculation or
the resummed version, Eq.~(\ref{xresum}) or Eq.~(\ref{bsgaminv}),
written as a function of the ``effective mass'' $m_b^* = m_b + k_+$,
{\it i.e.}, $x = 2E/m_b^* = M_Bx_B/m_b^*$.  The new differential rate
is now a function of $x_B=2E/M_B$.  The addition of the structure
function resums the leading-twist corrections and moves the endpoint
of the spectrum from $E=m_b/2$ to the physical endpoint
$E=M_B/2$.\footnote{The true end point takes into account the final
state masses, but these corrections are higher order in the heavy
quark mass expansion.}  The cut rate may then be written as
\begin{equation}
\label{cutratestruc}
\Gamma_H\left[\frac{2 E_{cut}}{M_B}\right] =
  \int_{2 E_{cut}-m_b}^{\bar{\Lambda}}dk_+\,f(k_+)\,\Gamma_p
  \left[\frac{2 E_{cut}}{m_b+k_+}\right]
\end{equation}
where $\Gamma_p [2 E_{cut}/(m_b+k_+)]$ is the partonic rate with a cut
at $x_p=2 E_{cut}/(m_b+k_+)$.

\section{Extraction of $|V_{\lowercase{ub}}|^2/
          |V_{\lowercase{ts}}^*V_{\lowercase{tb}}|^2$}

From the above discussion, it is clear that a model independent
extraction of $|V_{ub}|^2$, based solely on inclusive semi-leptonic
decays, will rely on a precise determinations of $\bar{\Lambda}$ and
$\lambda_1$, as well as the shape of the structure function $f(k_+)$.
Knowing the first few moments of $f(k_+)$, $\bar{\Lambda}$ and
$\lambda_1$ alone is perhaps not sufficient, since the shape of the
structure function is sensitive to higher moments as well.

Here we propose a method to extract the ratio $|V_{ub}|^2/|V_{ts}^*
V_{tb}|^2$ from the charged lepton spectrum of $B\to X_u\ell\bar{\nu}$
and the photon spectrum of $B\to X_s\gamma$, without invoking any
knowledge of $f(k_+)$. This method is based on the observation that
the only difference in the resummations between these two processes is
the $g_{s\ell}$ piece in the Sudakov exponent and the $x_\nu$ integral
in the semi-leptonic decays.  Applying Eq.~(\ref{intmell}) on
Eq.~(\ref{moment}), and using the notation introduced in
Eq.~(\ref{mellinnotation}), we get
\begin{eqnarray}
{\cal M}\left[\frac{d\Gamma}{dx};N\right]&=&
  \Gamma_0 \int dx_\nu \,6(1-x_\nu)x_\nu H(x_\nu) \;
  \frac1N\left( S_{N+1} \sigma_{N+1} J^\gamma_{N+1}\right)\;
  e^{g_{s\ell}(\chi',\, x_\nu)} \nonumber\\
 &=&  \Gamma_0 \int dx_\nu \,6(1-x_\nu)x_\nu H(x_\nu) \;
 \frac1{H^\gamma}\left(\frac{1}{\Gamma_0^\gamma}
  {\cal M}\left[x\frac{d\Gamma^\gamma}{dx};N\right]\right)
   \frac{e^{g_{s\ell}(\chi',\,x_\nu)}}N \;,
\end{eqnarray}
where
\begin{equation}
\chi' = \alpha_s \beta_0 \log(N+1) \approx \alpha_s \beta_0 \log(N).
\end{equation} 
Then taking the inverse Mellin transform of both sides and using the
Convolution Theorem Eq.~(\ref{conmell}) on the right hand side we get
\begin{equation}
\label{ratio}
\frac{d\Gamma}{dx} =  
   \frac{\Gamma_0}{\Gamma^\gamma_0\;H^\gamma} \int dx_\nu
   \,6 x_\nu (1-x_\nu)  H(x_\nu) \; \int_x^{M_B/m_b} \frac{du}{u}
   \,u^3\,\frac{d\Gamma^\gamma}{du}\,
   {\cal M}^{-1}\left[\frac{e^{g_{s\ell}(\chi,\,x_\nu)}}{N};\frac{x}{u}
   \right],
\end{equation}
where we have replaced $\chi'$ with $\chi$.  Let us make a few
comments on these results. First, $d\Gamma/dx$ and $d\Gamma^\gamma/du$
are both experimentally measurable quantities. We have managed to
relate them in terms of theoretically known functions.  The only
dependence on the non-perturbative quantities, $\bar{\Lambda}$ and
$\lambda_1$, resides in the upper limit, $M_B/m_b$, for the $u$
integral.  Furthermore, the dependence on the structure function has
been eliminated. A factor of $u^2$ has also been added so that we may
correctly reproduce the total rate to order of $\bar{\Lambda}/m_b$.
Powers of $u$ are subleading in $1/N$ and thus we have the freedom to
include this correction. However, we make no claims about reproducing
corrections of order $\alpha_s \bar{\Lambda}/m_b$ or
$(1-x)\bar{\Lambda}/m_b$.

Using the results in the Appendix the inverse Mellin transform in
Eq.~(\ref{ratio}) can be easily performed
\begin{equation}
{\cal M}^{-1}\left[ \frac{e^{g_{s\ell}(\chi,\,x_\nu)}}N;
  \frac{x}{u}\right] = \theta(1-x/u)
  e^{g_{s\ell}(\alpha_s\beta_0\, l_{x/u},\,x_\nu)}\;.
\end{equation}
Here $l_{x/u} = -\log[-\log(x/u)]$. Therefore
\begin{eqnarray}
\label{iniK}
\frac{d\Gamma}{dx} &=&  
 \frac{\Gamma_0}{\Gamma^\gamma_0} \int dx_\nu
\,6 x_\nu (1-x_\nu) \frac{H(x_\nu)}{H^\gamma}\; \int_x^{M_B/m_b} du\;
 u^2\,\frac{d\Gamma^\gamma}{du}\,
 e^{g_{s\ell}(\alpha_s\beta_0\,l_{x/u},\,x_\nu)} \nonumber\\
 &=& \frac{|V_{ub}|^2}{|V_{ts}^* V_{tb}|^2}
 \frac{\pi}{3\,\alpha\,C_7(m_b)^2}\,
   \int_x^{M_B/m_b} du\;u^2\,\frac{d\Gamma^\gamma}{du}\,
   K\left[x;\frac4{3\pi\beta_0}\log(1-\alpha_s\beta_0\,l_{x/u})\right],
\end{eqnarray}
where the function $K(x;y)$ in the convolution is the integral
over the neutrino energy $x_\nu$, taking into account the $x$
dependence up to ${\cal O}((1-x)^2)$, and is given by
\begin{eqnarray}
K(x;y) &=& 6\left\{\left[1+\frac{4\alpha_s}{3\pi}\left(1
 -  \psi^{(1)}(4+y) \right)\right]\frac1{(y+2)(y+3)}
  \right.\nonumber\\
&&\phantom{6\{}- \frac{\alpha_s}{3\pi}
  \left[\frac{1}{(y+2)^2}-\frac{7}{(y+3)^2}\right]
 -\left.\frac{4\alpha_s}{3\pi}\left[ \frac1{(y+2)^3}-\frac1{(y+3)^3}\right]
  \right\} \nonumber\\
&& - 3(1-x)^2.
\end{eqnarray}
In the above expression we have restored the $x$ dependence 
in the $x_\nu$ integral up to
${\cal O}((1-x)^2)$ and $\psi^{(1)}(z)$ is the first derivative of the
{\em digamma function}
\begin{equation}
\psi(z) = \frac1{\Gamma(z)}\frac{d}{dz}\,\Gamma(z).
\end{equation}
Using the perturbative one loop level prediction for the radiative
decay rate in Eq.~(\ref{iniK}) and expanding in $\alpha_s$ leads to
the one loop prediction for the semi-leptonic rate up to corrections
of order ${\cal O}(\alpha_s(1-x))$ and ${\cal O}((1-x)^3)$.

The Wilson coefficient $C_7(\mu)$ has been computed to next-to-leading
order accuracy in recent years. Here we quote from Ref.~\cite{CMM,KN}
its value at the scale $m_b$
\begin{equation}
C_7(m_b) = -0.31 + 0.48\, \frac{\alpha_s}{4\pi} +
        0.03\, \frac{\alpha}{\alpha_s}.
\end{equation}
More explicitly, $|V_{ub}|^2/|V_{ts}^*V_{tb}|^2$
may be extracted from the relation
\begin{eqnarray}
\label{forexp}
\frac{|V_{ub}|^2}{|V_{ts}^* V_{tb}|^2} &=&
\frac{3\,\alpha\,C_7(m_b)^2}{\pi (1+3\bar{\Lambda}/M_B) }
   \int^1_{x_B^c}dx_B\frac{d\Gamma}{dx_B}\nonumber\\
&&\times\,\left\{\int^1_{x_B^c}dx_B \int_{x_B}^{1} du_B\;u_B^2
   \,\frac{d\Gamma^\gamma}{du_B}\,
   K\left[x_B;\frac4{3\pi\beta_0}\log(1-\alpha_s\beta_0\,l_{x_B/u_B})\right]
   \right\}^{-1},
\end{eqnarray}
where we have rewritten our result in terms of the hadronically scaled
variable $x_B=2E/M_B$, $\bar{\Lambda}=M_B-m_b$ and the
experimental cut $x_B^c$. We may use the central value for
$\bar{\Lambda}$, $390{\rm\ MeV}$, with a $110{\rm\ MeV}$ standard
deviation\cite{GKLW}.\footnote{For consistency one should use the one
loop extracted value of $\bar{\Lambda}$.}

One might be concerned about the existence of the Landau pole in
$g_{s\ell}(\chi)$, which renders its inverse Mellin transform
imaginary before $x_B = 1$. In practice, this will not be a
problem. The Landau pole in $g_{s\ell}$ is located at $(x_B/u_B)_{max}
= 1 - \exp[-1/(\alpha_s\beta_0)] \approx 0.99913$, so numerically, one
can just integrate $u_B$ in Eq.~(\ref{forexp}) from $x_B/0.99$ to $1$
to avoid the Landau pole without incurring substantial error, since
the measured rate is a smooth function.

\section{Theoretical Uncertainties and Conclusions}

By relating two disparate rates near their respective endpoints we
have eliminated the uncertainty due to our ignorance of the structure
function. However, we are left with some unknown uncertainty due to
the assumption of quark-hadron duality. As usual, it is very difficult
to quantify the errors due to this assumption. These errors exist in
all predictions of heavy quark decay rates though qualitatively, we
expect them to be larger in some cases than in others. Shifman and
collaborators \cite{shif} have argued the errors due to rotating from
Euclidean space (this may be equated with the error in coming close to
the cut in the dispersion relation), where the OPE is performed, back
to Minkowski space leads to power suppressed corrections. They showed
this to be true in some models of QCD. We would expect the size of
this power correction to depend upon the size of the region over which
the differential rate is smeared. Heuristically, this just means that
the partonic rate should approach the average of the hadronic rate as
we smear over resonances.  In our case, we smear over a region of
invariant mass between $M_\pi$ and $M_D$, so we would expect smearing
over a significant number of final states, which in some sense is a
measure of the sensitivity to parton-hadron duality. However, the
spectrum is weighted towards smaller invariant mass states. Thus, if
the decay in the end point region only went to the one pion state, we
would expect significant errors due to duality. As the number of final
states sampled increases, we would expect the duality errors to
dwindle.

In addition to the duality errors, there will also be the usual errors
due to higher order corrections in $1/m_b$.  By including the factor
of $u^2$ in Eq.~(\ref{forexp}) we correctly reproduce the total rate up
to corrections which are down by $\Lambda/m_b$.  
Thus the leading non-perturbative
corrections, which have been left out, are of order $\Lambda^2/m_b^2$
and $(1-x)\Lambda/m_b$.  These corrections include the finite pion
mass effects which shift the endpoint from $M_B/2$ to
$(M_B^2-M_\pi^2)/(2M_B)$.  The error due to the
uncertainty in $\bar{\Lambda}$, $\delta\bar{\Lambda}$, which is
dominated by statistical errors, is
$3\,\delta\bar{\Lambda}/m_b\simeq5\%$ and will be reduced in the near
future.

Finally, there are the errors due to subleading terms in the exponent
of the resummed result Eq.~(\ref{resummed}). This issue may be
addressed by determining the effects of resummation which we have
included. In \cite{LR} it was found that to get a reliable prediction
for $B\rightarrow X_s\gamma$, given the present experimental cut of
$x_c=0.81$, it was not necessary to sum the endpoint logs. Is the same
true in the present case? We may estimate the effects of resummation
by using a model to make a prediction for the $d\Gamma^\gamma/du_B$ in
Eq.~(\ref{forexp}), and then compare the result for the cut rate with
the rate we would find had we expanded the $\exp(g_{sl})$ in powers of
$\alpha_s$.  Doing so we find that the effects of resummation are of
order ten percent.  Of course, as emphasized in \cite{LR}, the true
effect of resummation will depend upon the measured shape of the
radiative end point spectrum.  However, we do not expect that the
effect of the resummation to be so important that the inclusion of
$g_3$ would be relevant.

Before closing, we should point out that with experimental progress in
reconstructing the neutrino momentum, it may be feasible to make a cut
on the hadronic invariant mass instead of the lepton energy to reject
charmed final states \cite{dai}.  This technique, while experimentally
more difficult, has the advantage that the spectrum is not
preferentially weighted toward low invariant mass states.  However, as
in the case discussed in this paper, the rate will again be sensitive
to the Fermi motion of the $b$-quark \cite{FLW,MR,FN}.  Thus, an
approach analogous to the one developed here will be necessary to get
a reliable extraction of $|V_{ub}|$. This work will be reported in a
separate publication.

\acknowledgments 
This work was supported in part by the Department of Energy under
grant number DOE-ER-40682-143.

\appendix
\section*{} 
Here we derive the inverse Mellin transform Eq.~(\ref{bsgaminv}).  In
Ref.~\cite{CMNT} the inverse Mellin transform was calculated to
leading logarithm accuracy. For our purposes, we need to evaluate the
$\log (N)\,g_1[\alpha_s \log (N)]$ piece to next-to-leading logarithm
accuracy, whereas $g_2[\alpha_s \log(N)]$ itself is already of this
order. The Mellin transform pair in our case is defined as
\begin{eqnarray}
\label{mellinnotation}
  {\cal M}[f(x);N] &=& f^*(N) = \int_0^1\, x^{N-1} f(x)\, dx , \nonumber\\
  {\cal M}^{-1}[f^*(N);x] &=& f(x) = \frac1{2\pi i}
  \int_{C-i\infty}^{C+i\infty} x^{-N} f^*(N)\,dN,
\end{eqnarray} 
We will also make use of the following two formulae which can be
easily derived
\begin{eqnarray}
\label{intmell}
{\cal M}\left[x f(x); N\right] &=& f^*(N+1) \;, \\
\label{conmell}
{\cal M}^{-1} \left[f^*(N)\, g^*(N) ; x\right] &=&
   \int_x^1 f\left(\frac{x}{u}\right)\,g(u) \frac{du}{u} \;.
\end{eqnarray}

To evaluate the inverse Mellin transform, let's begin by considering
the following integral
\begin{eqnarray}
\int_{C-i\infty}^{C+i\infty}\,\frac{dN}{N}\, x^{-N}
  e^{\log (N)\,F_1[\alpha_s \log (N)]} &=&
\int_{C'-i\infty}^{C'+i\infty}\,\frac{dz}{z} \, e^{z +
(\log z + l)\,F_1[\alpha_s(\log z + l)]} \nonumber\\
&=& e^{l\;F_1(\alpha_s\, l)}\times\int\frac{dz}{z} \,
e^{z + G(\log z,\alpha_s,l)}\;,
\end{eqnarray}
where  $z=-N \log x$ and
\begin{eqnarray}
\label{gfunc}
G(\log z,\alpha_s,l) =&& (\log z + l)\,F_1[\alpha_s(\log z + l)]
                      - l\;F_1[\alpha_s\, l] \nonumber\\
=&& \log z\left( F_1[\alpha_s\,l] + \alpha_s l\, F_1'[\alpha_s\,l]\right)
    \nonumber\\
&+& \log^2 z \left(\alpha_s F_1'[\alpha_s\,l]+\frac1{2}\alpha_s
     (\alpha_s\,l) F_1''[\alpha_s\,l] \right) \nonumber\\
&+& \cdots\; ,\\
l = &-&\log(-\log x)\;,\\
 C'= &-&C \log x \;.
\end{eqnarray}
In writing down the above we have assumed that $\alpha_s\;l$ is ${\cal
O} (1)$ and thus Taylor-expand with respect to $\alpha_s\log z$. The
coefficient of $\log z$ in Eq.~(\ref{gfunc}) represents contributions
to next-to-leading logarithm of the form $\alpha_s^n\,l^n$. Terms with
$\log^n z$, $n \geq 2$ only contribute to next-to-next-to-leading
logarithm as can be seen by expanding its exponential and integrating
term by term in $z$. We get precisely an expression of the form $1 +
{\cal F}(\alpha_s,l)$ as in Eq.~(\ref{invmell}). In general, if we
want to perform the inverse Mellin transform to the order of
$\alpha_s^{n+k}\,l^n$, we need to keep $G$ up to terms of $\log^{k+1}
z$ and evaluate the contour integral.  However, we would have to
include in Eq.~(\ref{sudseries}) higher order contributions, such as
$\alpha_s g_3$, for consistency.

Truncating $G$ at $\log z$ order and evaluating the integral we then
have
\begin{eqnarray}
\frac1{2 \pi i}\; \int_{C'-i\infty}^{C'+i\infty}\,
 \frac{dz}{z}\; e^{z}\, z^{F_1(\alpha_s\,l) + \alpha_s l\, F_1'(\alpha_s\,l)}
&=&
  \frac1{2 \pi}\;   \int_{-\infty}^\infty dy\; e^{C'+ i y} \,
(C'+ i y)^{F_1(\alpha_s\,l) + \alpha_s l\, F_1'(\alpha_s\,l)-1}
 \nonumber\\
&=& \frac1
{\Gamma\left[1 - F_1(\alpha_s\,l) - \alpha_s l\, F_1'(\alpha_s\,l)\right]}\,,
\end{eqnarray}
where we have changed variable from $z$ to $z = C' + i y$ and have
used
\begin{equation}
\label{laplace}
\frac1{\Gamma(z)} = \frac1{2 \pi} \int_{-\infty}^\infty du\;
                    e^{a + i u}\; (a + i u)^{-z}.
\end{equation}
Including the $F_2$ piece we have
\begin{eqnarray}
\label{inv}
&& \frac{1}{2\pi i} \int_{C-i\infty}^{C+i\infty}\,\frac{dN}{N}\,x^{-N}\,
  e^{\log(N)\,F_1[\alpha_s \log(N)]+ F_2[\alpha_s \log(N)]} \nonumber\\
&&
\phantom{\frac{1}{2\pi i} \int_{C-i\infty}^{C+i\infty}\,\frac{dN}{N}\,x^{-N}}=
   \theta(1-x) \frac{e^{l\:F_1(\alpha_s l) + F_2(\alpha_s l)}}
   {\Gamma\left[1 - F_1(\alpha_s l) - \alpha_s l \,F_1'(\alpha_s l)\right]}
   \times \left[1 + {\cal F}(\alpha_s ,l) \right],
\end{eqnarray}
where
\begin{equation}
{\cal F}(\alpha_s ,l) = \sum_{k=1}^{\infty} \alpha_s^k \,
                      \sum_{j=0}^{k-1} f_{kj}\,l^j
\end{equation}
represents next-to-next-to-leading log contributions.

By taking a derivative of both sides of Eq.~(\ref{inv}) with respect to
$\log x$, we arrive at
\begin{eqnarray}
\label{srinvmell}
&& \frac1{2 \pi i}\;\int_{C-i\infty}^{C+i\infty}\,dN\,
  e^{-N \log (x)+ \log (N)\,F_1[\alpha_s \log (N)] +
  F_2[\alpha_s \log (N)] } \nonumber\\
&&\phantom{\frac1{2 \pi i}\;\int_{C-i\infty}^{C+i\infty}}=
  -x \frac{d}{dx}\;\left\{\theta(1-x) \frac{e^{l\:F_1(\alpha_s l) +
  F_2(\alpha_s l)}}
  {\Gamma\left[1 - F_1(\alpha_s l) - \alpha_s l \,F_1'(\alpha_s l)\right]}
    \times \left[1 + {\cal F}(\alpha_s ,l) \right] \right\},
\end{eqnarray}
which is relevant in processes like $B \rightarrow X_s\gamma$ and
Drell-Yan production.

In the inclusive semi-leptonic decays we need to evaluate, instead,
the following integral
\begin{eqnarray}
\label{exact}
&& \frac{1}{2\pi i} \int_{C-i\infty}^{C+i\infty}\,\frac{dN}{N}\,x^{-N}\,
e^{\log(N+1)\,F_1[\alpha_s \log(N+1)]} \nonumber\\
&&
\phantom{\frac{1}{2\pi i} \int_{C-i\infty}^{C+i\infty}\,\frac{dN}{N}\,x^{-N}}=
   \frac1{2\pi i} \int \, \frac{dN'}{N'-1} \, x^{1-N'} \,
     e^{\log(N')\, F_1[\alpha_s \log(N')]} \nonumber\\
&&
\phantom{\frac{1}{2\pi i} \int_{C-i\infty}^{C+i\infty}\,\frac{dN}{N}\,x^{-N}}=
   x\, \sum_{k=0}^{\infty} \; \frac1{2\pi i} \int\,\frac{dN'}{N'}\,
 \left(\frac1{N'}\right)^k\,x^{-N'}\,e^{\log(N')\, F_1[\alpha_s \log(N')]} 
        \nonumber \\
&&
\phantom{\frac{1}{2\pi i} \int_{C-i\infty}^{C+i\infty}\,\frac{dN}{N}\,x^{-N}}=
    x\, e^{l\,F_1(\alpha_s\,l)} \, \sum_{k=0}^{\infty}\,(-\log x)^k \;
   \frac1{2\pi i} \int \frac{dz}{z}\,z^{-k} \, e^{z + G(\log z,\alpha_s,l)} \, .
\end{eqnarray}
Apply Eq.~(\ref{laplace}) and use the following sum
\begin{equation}
\sum_{k=0}^{\infty} \frac{(-L)^k}{\Gamma(1-F+k)} = (-L)^F e^{-L}
   \left[ 1 - \frac{\Gamma(-F,-L)}{\Gamma(-F)} \right],
\end{equation}
where $\Gamma(a,z)$ is the {\it incomplete Gamma function} defined as
\begin{equation}
\Gamma(a,z) = \int_z^\infty \, t^{a-1}\,e^{-t}\, dt,
\end{equation}
we then obtain
\begin{eqnarray}
&&\frac{1}{2\pi i} \int_{C-i\infty}^{C+i\infty}\frac{dN}{N}x^{-N}\,
e^{\log(N+1)\,F_1[\alpha_s \log(N+1)]} \nonumber \\
&&\phantom{\frac{1}{2\pi i} \int_{C-i\infty}^{C+i\infty}}=
  \theta(1-x)\,x\, e^{l\,F_1(\alpha_s \, l)}\;\sum_{k=0}^{\infty}
  \frac{(-\log x)^k}
  {\Gamma\left[1-F_1(\alpha_s l)-(\alpha_s l)F_1'(\alpha_s l)+k\right]}
   \nonumber\\
&&\phantom{\frac{1}{2\pi i} \int_{C-i\infty}^{C+i\infty}}=
  \theta(1-x)\, e^{l\,F_1(\alpha_s \, l)}\,
  \left(-\log x\right)^{F_1 + \alpha_s l\,F_1'}
  \left[1-\frac{\Gamma(-F_1-\alpha_s l F_1', -\log x)}
  {\Gamma(-F_1-\alpha_s l F_1')}\right],
\end{eqnarray}
where we have omitted the next-to-leading log contributions coming
from $F_2$, which would give an overall factor of $\exp[F_2(\alpha_s
l)]$.

On a practical note we find that, to the accuracy we are working, it
is simpler to use large $N$ approximation on the exponent of
Eq.~(\ref{exact}), {\em i.e.},
\begin{eqnarray}
\label{largen}
\frac{1}{2\pi i} \int_{C-i\infty}^{C+i\infty}\,\frac{dN}{N}\,x^{-N}\,
   e^{\log(N+1)\,F_1[\alpha_s \log(N+1)]} &\approx&
   \frac{1}{2\pi i}
   \int_{C-i\infty}^{C+i\infty}\,\frac{dN}{N}\,x^{-N}\,
   e^{\log (N)\,F_1[\alpha_s \log(N)]} \nonumber\\
&=& \theta(1-x)
   \frac{e^{l\,F_1(\alpha_s l)}}{\Gamma\left[1 - F_1(\alpha_s l)
     -\alpha_s l F_1'(\alpha_s l)\right]}.
\end{eqnarray}

\tighten

} 


\begin{references}

\bibitem{BA} 
For a review see {\it The BABAR Physics Book: Physics at an Asymmetric
B Factory}, ed. P. F. Harrison and H. R. Quinn, SLAC-R-0504, (1998).

\bibitem{MW}
A.V. Manohar and M.B. Wise, {\it Heavy Quark Physics}, Cambridge
University Press, in press.

\bibitem{ligeti} For a recent review on methods for extracting
$V_{ub}$ along with reasonable assessments of theory errors, see
Z. Ligeti, hep-ph/9908432.

\bibitem{CGG}
J. Chay, H. Georgi, and B. Grinstein, Phys. Lett. {\bf B247} (1990) 399;
M. Voloshin and M. Shifman, Sov. J. Nucl. Phys. {\bf 41} (1985) 120.

\bibitem{BSUV} 
I.I. Bigi, M.A. Shifman, N.G. Uraltsev, and A.I. Vainshtein,
Int. J. Mod. Phys. {\bf A9} (1994) 2467.

\bibitem{N}
M. Neubert, Phys. Rev. {\bf D49} (1994) 3392.

\bibitem{MN}
T. Mannel and M. Neubert, Phys. Rev. {\bf D50} (1994) 2037.

\bibitem{buras} For a review on calculating the
effective weak Hamiltonian, see A. Buras hep-ph/9901409.

\bibitem{ONEL}
M. Je\.{z}abek and J.H. K\"{u}hn, Nucl. Phys. {\bf B320} (1989) 961;
A.V. Manohar and M.B. Wise, Phys. Rev. {\bf D49} (1994) 1310; I.I. Bigi,
M. Shifman, N.G. Uraltsev and A.I. Vainshtein, Phys. Rev. Lett. {\bf 71}
(1993), 496; B. Blok, L. Koyrakh, M. Shifman and A.I. Vainshtein, Phys. Rev.
{\bf D49} (1994), 3356; ERRATUM-ibid. {\bf D50} (1994) 3572; T. Mannel,
Nucl. Phys. {\bf B413} (1994) 396.

\bibitem{LR}
A.K. Leibovich and I.Z. Rothstein, hep-ph/9907391.

\bibitem{SK}
G. Sterman and G.P. Korchemsky, Phys. Lett. {\bf B340} (1994) 96.

\bibitem{AR}
R. Akhoury and I.Z. Rothstein, Phys. Rev. {\bf D54} (1996) 2349.

\bibitem{CMNT} 
S. Catani, M.L. Mangano, P. Nason, and L. Trentadue,
Nucl. Phys. {\bf B478} (1996) 273.

\bibitem{AG}
A. Ali and C. Greub, Z. Phys. {\bf C49} (1991) 431; 
Phys. Lett. {\bf B259} (1991) 182;
Phys. Lett. {\bf B287} (1992) 191.

\bibitem{CLEO}
S. Glenn {\it el al.}, CLEO Collaboration, CLEO CONF 98-17.

\bibitem{GKLW}
M. Gremm, A. Kapustin, Z. Ligeti, and M.B. Wise,
Phys. Rev. Lett. {\bf 77} (1996) 20.

\bibitem{DSU}
R.D. Dikeman, M. Shifman, and N.G. Uraltsev,
Int. J. Mod. Phys. {\bf A11} (1996) 571.

\bibitem{CMM}
K. Chetyrkin, M. Misiak, and M. Munz, Phys. Lett. {\bf B400} (1997)
206; Erratum-ibid. {\bf B425} (1998) 414.

\bibitem{KN}
A. Kagan and M. Neubert, Euro. Phys. J., {\bf C7} (1999) 5.

\bibitem{shif}
M. Shifman, hep-ph/9405256; hep-ph/9505289;
B. Chibisov, R. Dikeman, M. Shifman and N. Uraltsev,
Int. J. Mod. Phys.
{\bf A12} (1997) 2075.

\bibitem{dai} J. Dai, Phys. Lett. {\bf B333} (1994) 212.

\bibitem{FLW} A. Falk, Z. Ligeti and M.B. Wise, Phys. Lett. {\bf
406} (1997) 225.

\bibitem{MR} T. Mannel and S. Recksiegel, Phys. Rev. {\bf D60}
 (1999) 114040.

\bibitem{FN} F. De Fazio and M. Neubert, JHEP 9906:017, 1999.

\end{references}
\end{document}